\documentclass{article}

\title{Bayesian Design and Analysis of Precision Trials with Partial Borrowing}
\pdfoutput=1
\date{}
\usepackage{graphicx} 
\usepackage{amsmath,latexsym, graphics,graphicx,showexpl,amsthm,amsfonts,fullpage}
\usepackage[left=1in,top=1in,right=1in,nohead]{geometry}
\usepackage{graphicx}
\usepackage[margin=10pt,font={small},labelfont=bf]{caption}
\usepackage{subcaption}
\usepackage{bbm} 
\usepackage{setspace}
\usepackage{natbib}
\usepackage[titletoc]{appendix}
\usepackage{epstopdf}
\usepackage{pifont}
\usepackage[pdftex,bookmarks,colorlinks=true,linkcolor=blue]{hyperref}
\usepackage{algpseudocode,algorithm}
\usepackage{authblk}
\usepackage{multicol}
\usepackage{hyperref}
\usepackage{array}
\usepackage{etoolbox}
\usepackage{setspace}
\usepackage{arydshln}
\usepackage{comment}
\usepackage[table]{xcolor}
\usepackage{multirow}

\newcolumntype{L}[1]{>{\raggedright\let\newline\\\arraybackslash\hspace{0pt}}m{#1}}
\newcolumntype{C}[1]{>{\centering\let\newline\\\arraybackslash\hspace{0pt}}m{#1}}
\newcolumntype{R}[1]{>{\raggedleft\let\newline\\\arraybackslash\hspace{0pt}}m{#1}}
\AtBeginEnvironment{thebibliography}{\linespread{1.7}\selectfont}
\makeatother
\author[1]{Shirin Golchi\thanks{e-mail: shirin.golchi@mcgill.ca}}
\author[2]{Satoshi Morita}

\affil[1]{McGill University, Department of Epidemiology, Biostatistics and Occupational Health}
\affil[2]{Kyoto University, Department of Biomedical Statistics and Bioinformatics}
\begin{document}

\maketitle
\begin{abstract}
With the advancement of precision medicine there is an increasing need for design and analysis methods in clinical trials with the objective of investigating effect heterogeneity and estimating subgroup effects. As this requires precise estimation of interaction effects, borrowing information from external data sources including retrospective studies and early phase clinical trials to enrich the trial in sparse subgroups is pertinent. Motivated by a trial in gastric cancer we consider a practical design and analysis framework for borrowing from external data sources that only partially inform the inference. As the analysis model we propose an individually weighted model where the external data are weighted based on their fit with the target population based on the distribution of a set of covariates. In a simulation study we assess the performance of the model under various scenarios and make comparisons to dynamic borrowing. In addition, we provide a Bayesian design framework where design priors are extracted from the external data to determine decision boundaries and sample sizes. The design procedure is demonstrated within the context of our motivating example.
\end{abstract}

\noindent%
{\it Keywords: Bayesian operating characteristics; Individualized weights; External data; Power prior; Precision medicine; Subgroup effects.
}

\section{Introduction}
\label{sec:intro}

Utilizing historical or external data to strengthen statistical inference in clinical trials has been explored in a variety of settings \citep{DemSelWee1981, Tar1982, IbrRyaChe1998}. Information borrowing may be of interest for a variety of reasons including power gains in presence of sample size constraints in the context of rare diseases or pediatric trials  \citep{zocholl_informing_2026}, constructing or complementing the control group in single arm trials or trials with imbalanced randomization ratios in oncology and early phase drug development \citep{ThaSim90, ZohTerZho08, CheCheMor16}. Depending on the objectives of the study, trial data may be augmented in one or more treatment groups. 

The statistical literature on external borrowing in clinical trials spans a spectrum of methods with the common objective of detecting and adjusting for the biases that may arise from discordance in information obtained from different sources. Particularly, several Bayesian methods have been proposed as the Bayesian framework offers a formal way of incorporating information via a prior distribution \citep{NeuCapBra2010, SchGstRoy14, RovFri19, LewSarZhu19, WebLiSea19, hobbs_hierarchical_2011, jiang_elastic_2023}. 
Among the existing methods, power priors \citep{CheIbr2000}, which discount external information by raising the likelihood constructed from the external data to a parameter, have enjoyed increased popularity due to their flexibility and generalizeability. The role of the power (or weight) parameter is to control the level of contribution of external information to inference and it may be specified in a static or dynamic manner \citep{IbrCheGwo15,duan2005modified,duan2006evaluating,carvalho2021normalized}. While The original formulation of the power prior relies on study level discounting, recent developments have extended the framework to accommodate partial borrowing \citep{Lu02012022, kwiatkowski_case_2024, alt_leap_2024}. Noteably, \cite{alt_leap_2024} propose the latent exchangeability prior (LEAP) which assigns a power parameter to the individual level likelihoods. The individual level powers are then dynamically inferred from the data in a fully Bayesian model.

With the advancement of precision medicine which aims to adapt treatment recommendations based on individual patient characteristics or biomarkers, inferring effect heterogeneity across subgroups and estimating subgroup specific effects is of primary or secondary interest in many clinical trials. Detecting effect heterogeneity requires estimating treatment-covariate interaction effects with sufficient precision. As small subgroup sizes are common in clinical trials, most trials are underpowered with respect to detection of effect heterogeneity and subgroup analysis. Therefore, borrowing information from external sources is particularly pertinent in the context of precision trials investigating effect heterogeneity and subgroup effects. 

While most of the work on external borrowing is focused on average treatment effects, recent developments have aimed to address subgroup analyses using external information. \cite{daniells_incorporating_2025} consider a basket trial design where information is borrowed across baskets and from external sources via the exchangeablity-nonexchangeablity (EXNEX) model that incorporates a power prior in the non-exchangeability part of the mixture model. \cite{schwartz_harmonized_2025} propose harmonized estimator for the subgroup specific effects that borrow from external control data but remains coherent with the average treatment effect estimate based on the current trial data.

In this manuscript we consider information borrowing toward subgroup effects from external data sources that may inform differing subsets and functions of the model parameters. We propose an individually weighted formulation of the power prior where the power parameters are specified using a similarity measure inspired by the double-dipper similarity function proposed by \cite{PageQuintana2018} based on the available covariates. We propose truncating the weights distribution to discard external data that score low according to this similarity measure where the truncation cut-off may be specified to limit the effective sample size of the external data. We assess the performance of the individually weighted model under a rage of simulation scenarios and make comparisons with LEAP as a representative of dynamic borrowing methods. 

In addition, we propose a Bayesian design procedure that builds upon the method of \cite{psioda_bayesian_2019} to specify design priors but accommodates the incomplete structure of the external data. We showcase the analysis and design framework with external borrowing within the context of a phase II randomized clinical trial in gastric cancer where retrospective and single arm external data are available to inform the treatment effect on the subgroup of patients with recurrent disease status.



The remainder of this article is organized as follows. The gastric cancer trial example is presented as our motivating example in Section~\ref{Sec:me}. The individually weighted model for analysis with external borrowing is described in Section~\ref{sec:methods} within a precision trial framework from analysis and design perspectives. Section~\ref{sec:sim} follows with a simulation study to assess the model performance and make comparisons to dynamic borrowing. The proposed design and analysis methods are showcased in Section \ref{sec:app} in the context of the motivating example and a discussion follows in Section \ref{sec:dis}.

\section{Motivating Example}\label{Sec:me}
The motivating clinical trial for the present work is a Randomized phase II study of S-1 plus cisplatin (SP) versus capecitabine plus cisplatin (XP) in patients with advanced or recurrent gastric cancer that cannot be curatively resected (XParTS-II) \citep{XParTSII}. We consider the research question concerning the comparative effect of SP versus XP in the subgroup of patients with recurrent gastric cancer. However, this subgroup constitutes only 23\% of patients enrolled in the trial and the small sample size does not allow for sufficient power for detecting treatment effect within this subgroup. To address this challenge we aim to use data from external sources to the trial to gain power toward the subgroup specific treatment effect estimation. In particular, data are available from a Phase II study to evaluate the efficacy and safety of XP in patients with recurrent gastric cancer (XParTS-I) that can inform the XP arm parameter within the target subgroup \citep{XParTSI}. Data are also available from a retrospective study on the efficacy of SP in patients with recurrent gastric cancer which may be informative toward the subgroup specific parameter of the SP arm \citep{Retro}.

While we are interested to borrow information from both external sources, we would like to adjust for any discrepancies between the patient populations in these studies and the current trial. In the following section we propose to calculate a similarity measure for each patient data in the XParTS-I trial and the retrospective study compared to the XParTS-II trial patient population based on a set of available covariates. These similarity measures are then used to adjust the contribution of each external patient data to the analysis allowing us to take advantage of any overlap between the patient populations across the studies.

Moreover, we consider the hypothetical re-design of XParTS-II with the objective of estimating the subgroup specific effect while utilizing the XParTS-I and retrospective study data not only via the analysis prior but also to construct design priors. As the objective of this design exercise we investigate how partial borrowing from the external data sources can save sample size for the planned study.

\section{A precision trial framework}
\label{sec:methods}
Consider a randomized clinical trial (RCT) where the primary objective is to investigate effect heterogeneity and establish treatment effects within specific patient subgroups where the subgroup specific quantity of interest  (estimand) is denoted by $\Gamma$. Suppose that the analysis model is indexed by a set of parameters $\boldsymbol{\theta}$ and characterized by a likelihood function $f(\boldsymbol{\theta}; \mathbf{D})=\prod_{n=1}^N f(\boldsymbol{\theta}; \mathbf{d}_n)$ where $\mathbf{D}=\{\mathbf{d}_1, \ldots, \mathbf{d}_N\}$ represents the data collected through the trial and $\mathbf{d}_n$ is the data on patient $n = 1, \ldots, N$ which includes the outcome(s), arm assignment indicator, prognostic covariates, effect modifiers, censoring indicator or any other relevant variables. 

Particularly, we consider the following generalized linear model where the outcome of patient $n$, is modelled as
\begin{align}
\label{eq:AM}
&Y_n\sim f(y\mid\mathbf{x}_n, a_n, \mathbf{s}_n ; \boldsymbol{\theta}) \nonumber\\
&g(\text{E}(\mathbf{Y_n})) =  \eta(\mathbf{x}_n;\boldsymbol{\beta}) + \gamma( \mathbf{s}_n;\boldsymbol{\psi} )a_n, 
\end{align}
where $\boldsymbol{\theta} = (\boldsymbol{\beta}, \boldsymbol{\psi})$ are the model parameters. The vector $\mathbf{x}_n$ includes a set of (prognostic only or prognostic and effect modifier) covariates and $\mathbf{s}_n$ is a subset of covariates in $\mathbf{x}_n$ that are potential effect modifiers. The indicator $a_n$ specifies the treatment group assignment for patient $n$. The functions $\eta$ and $\gamma$ specify the relationship with prognostic and predictive covariates and can take simple linear or more flexible forms such as splines. The function $\gamma$ and its parameters explain how a patient with covariate values $\mathbf{s}_n$ responds to the new treatment.

Given the above formulation, the subgroup of patients for whom the treatment has a meaningful effect is defined by the following set,
\begin{equation}
\mathcal{S}^*=\{\mathbf{s}: P(\gamma( \mathbf{s};\boldsymbol{\psi} )>\delta_c\mid \mathbf{D})>(1-\epsilon)\},
\end{equation}
where $\delta_c$ is a minimum clinically meaningful effect and $\epsilon$ is an upper probability threshold that is tuned as a design parameter to meet desired operating characteristics.

We define the target of inference as the marginal treatment effect over the sensitive set which is obtained by integrating over this set with respect to the joint empirical distribution of covariates,
\begin{equation}
\Gamma(\boldsymbol{\theta}) = \int_{\mathcal{X}}\int_{\mathcal{S}^*} \left[E(\mathbf{Y}\mid \mathbf{x}, \mathbf{s}, a=1;\boldsymbol{\theta}) - E(\mathbf{Y}\mid \mathbf{x}, \mathbf{s}, a=0;\boldsymbol{\theta})\right]  \hat{p}(\mathbf{x},\mathbf{s})d\mathbf{x}d\mathbf{s}.
\end{equation}

Under a Bayesian analysis framework, inference is carried out by specifying an analysis prior for $\boldsymbol{\theta}$, $\pi_a(\boldsymbol{\theta})$ and obtaining the posterior distribution $\pi(\boldsymbol{\theta}\mid \mathbf{D})$. The posterior distribution of $\Gamma$, $\pi(\Gamma(\boldsymbol{\theta})\mid \mathbf{D})$ is then derived whose functionals are used for decision making. For example, efficacy or futility conclusions are made by comparing the a tail probability of $\pi(\Gamma(\boldsymbol{\theta})\mid \mathbf{D})$, to a prespecified threshold. 

We take interest in the particular scenario where due to feasibility constraints $\mathbf{D}$ does not provide sufficient information toward $\Gamma$. This can be because the sample size of the target subgroup is small or that $\mathbf{D}$ is sparse with respect to the respective effect modifier. In this case, we wish to use information from sources external to the trial to gain precision in estimating $\Gamma$ and power to make decisions according to the study hypotheses. In the following, we present an individual weighting scheme that adjusts the contribution of each external patient data according to their fit with the target population based on a set of covariates. 



\subsection{Individual discounting}\label{sec:IW}

 Denote the current clinical trial data set by $\mathbf{D}_I = (\mathbf{X}_I, \mathbf{a}_I, \mathbf{y}_I)$ and the external data by $\mathbf{D}_{E} = (\mathbf{X}_E, \mathbf{a}_E, \mathbf{y}_E)$ where $\mathbf{X}_I$ and $\mathbf{X}_E$ are the design matrices constructed from all covariates that are at least partially observed in both $\mathbf{D}_I$ and $\mathbf{D}_{E}$. The vector $\mathbf{y}$ is the vector of outcomes and $\mathbf{a}$ represents the vector of treatment group assignments. Note that the covariate and treatment distribution may well vary across different sources that constitute the external data set. For example, the external data may be partly control data where $\mathbf{a}=\mathbf{0}$ or that certain levels of categorical covariates be unobserved in a subset of the external data.



To specify $\pi_a(\boldsymbol{\theta})$, we use the following ``individually weighted'' prior to incorporate information from external data into the analysis,
\begin{equation}
\pi_a(\boldsymbol{\theta})=\pi_{IW}(\boldsymbol{\theta}  \mid \mathbf{D}_E) \propto \pi_0(\boldsymbol{\theta} )\left\{\prod_{n=1}^{N_E}f(\boldsymbol{\theta}; \mathbf{d}_n)^{\omega_{n}} \right\}
\end{equation}
where $\pi_0(\boldsymbol{\theta})$ is a weakly informative default prior that would be used in absence of any external data, and $\omega_{n}$ is a measure of similarity or fit for external patient $n$ to the current trial patient population and determines the contribution of patient $n$ to the prior. Combined with the likelihood of the internal data, this will give rise to the posterior distribution,
\begin{equation}
\label{eq:wIPD}
\pi_{IW}(\boldsymbol{\theta}  \mid \mathbf{D}_E, \mathbf{D}_I) \propto \pi_0(\boldsymbol{\theta} )\left\{\prod_{n=1}^{N_E}f(\boldsymbol{\theta}; \mathbf{d}_n)^{\omega_{n}} \right\}f(\boldsymbol{\theta}; \mathbf{D}_I).
\end{equation}

\subsection{Specification of the weights} \label{sec:weights}
 The weights represent a measure of similarity or fit between each external patient data and the internal data. This similarity is measured with respect to a set of prognostic covariates. Let us denote the matrix formed by these covariates by $\mathbf{Z}$ and the vector of ``weighting covariates'' for patient $n$ by $\mathbf{z}_n$. Note that $\mathbf{z}_n$ and $\mathbf{s}_n$ are vectors of covariates that are both subsets of $\mathbf{x}_n$ but do not overlap. Our goal is to borrow external data toward the enrichment of sparse patient subgroups defined by $\mathbf{S}$. However, weighting external data according to any variables in $\mathbf{S}$ would further down-weight external data that belong to parts of the distribution of $\mathbf{s}$ that are sparse within the trial. Therefore, based on our empirical results (see Supplementary Material), we recommend that in general the weighting matrix $\mathbf{Z}$ be formed by all prognostic-only covariates. 
 
 Various distance/discordance and similarity/agreement measures have been proposed either in weighting schemes similar to the model presented above or in other related contexts. For example, the weights may be specified as a propensity score estimated from a propensity model. \cite{PageQuintana2018} propose a variety of similarity measures within the context of covariate-informed clustering in product partition models.  

Here, we use the posterior predictive similarity function which is based on a \emph{similarity model}, $q$, that is defined to measure the closeness of the external distribution of covariates to that of the internal data. The weights can then be specified as the posterior predictive probability of each subject's ``fit'' given the internal data,
\begin{equation}
\label{eq:dd}
\omega_n = \int q(\mathbf{z}_n\mid \boldsymbol{\xi})\pi(\boldsymbol{\xi}\mid \mathbf{Z}_I)d\boldsymbol{\xi},
\end{equation}
where $\boldsymbol{\xi}$ is the vector of parameters indexing the similarity model and $\mathbf{Z}_I$ is the internal weighting covariate matrix. This measure is an extension of the auxiliary similarity function which was originally proposed by \cite{Muller_ppm_2011} and used a ``prior'' $\pi(\xi)$ instead of the posterior $\pi(\xi\mid \mathbf{Z}_I)$. 
As explained by \cite{PageQuintana2018}, the model $q$ need not be a joint probability distribution for the covariates but any function that generates larger values for more similar covariate vectors would be suitable. Therefore, we follow recommendations in \cite{Muller_ppm_2011} to define $q$ as a product of probability distribution functions for each covaraite based on the covariate type. For continuous covariates, we use kernel density estimates to capture bimodality and/or skewness. For binary and categorical variables we use the binomial and multinormial distributions, respectively.

In Section 1 of the Supplementary Material, we make comparisons with propensity score and the Gower similarity measure \citep{Gower1971} as alternative approaches for specifying the weights in a simulation study. Although the simulation study results favor the posterior predictive similarity function, we note that this is not an exhaustive investigation and other suitable similarity measures may be used in practice.

\subsection{Truncation of the weights}
The weighting scheme described above aims to mitigate potential biases that arise from incorporating external data under two conditions; confounders should be included in the covariate set that is used to compute the weights and the external data sets should be comparable in size or smaller than the trial data. If the external data are dominant in size, as may be the case for real world data, weighting may not be sufficient to protect against bias since inclusion of many participants with small weights can still have a significant contribution to the likelihood. In this case we propose discarding the portion of external data corresponding to the left tail of the distribution of weights using a cut-off value. This cut-off may be specified as a fixed quantile or such that the contribution of the external data with respect to the effective sample size (measured as the sum of $\omega_n$) remains less than or equal to that of the trial data, i.e., obtain the cut-off $\omega_0$ such that,
\begin{equation}
\label{eq:cutoff}
\sum_{n= 1}^{N_E}\omega_n\mathrm{I}(\omega_n>\omega_0)\leq N_I,
\end{equation}
where $\mathrm{I}(\cdot)$ is an indicator function. We acknowledge that this is a simplistic definition for the effective sample size as it assumes equal contribution for each external subject. However, this simplified measure of information serves sufficiently well for the purpose of specifying an approximate truncation limit.

\subsection{Design}
\label{sec:des}
We consider the above precision trial and data structure from a design perspective. In particular, we discuss how clinical trials may be designed to optimize power toward establishing subgroup effects by using external data sources that provide partial information on the model parameters. We present the research objectives in form of the following hypotheses,
\begin{equation}
H_0: \Gamma(\boldsymbol{\theta}) \notin (\Gamma_l, \Gamma_u) \hskip 20pt \text{vs} \hskip 20pt H_1:\Gamma(\boldsymbol{\theta}) \in (\Gamma_l, \Gamma_u)
\end{equation}
where $-\infty<\Gamma_l<\Gamma_u<\infty$. This formulation includes interval hypotheses such as equivalence, one-sided superiority and non-inferiority hypotheses. A decision rule may then be defined as rejecting the null hypothesis if $\tau(\mathbf{D})=P(\Gamma \in (\Gamma_l, \Gamma_u)\mid \mathbf{D})>\nu$ with the decision boundary $\nu$ specified at the design stage and $\mathbf{D} = (\mathbf{D}_E, \mathbf{D}_I)$.

We take a Bayesian approach toward design and use Bayesian operating characteristics for calibrating the decision rules and determine the sample size. The essential component of a Bayesian design, is a design (or sampling) prior distribution \citep{ohagan_bayesian_2001, gelfand_simulation-based_2002, best_beyond_2025}, distinct from the analysis prior, which is used to reflect the researcher's beliefs with appropriate level of uncertainty at the design stage. 

We consider constructing the design prior from the available data sources. As mentioned earlier we allow for the possibility that these data sources inform different subsets of the model parameters. In other words, with the external data alone, the analysis model for the planned trial needs to be identifiable only with respect to the target of inference $\Gamma$ and not with respect to all parameters. Following \cite{psioda_bayesian_2019} we define default null and alternative design priors, respectively, as
\begin{equation}
\pi_d^0(\boldsymbol{\theta})=\pi(\boldsymbol{\theta}\mid \mathbf{D}_E, \Gamma\notin (\Gamma_l, \Gamma_u)) \hskip 20pt \text{and} \hskip 20pt \pi_d^1(\boldsymbol{\theta})=\pi(\boldsymbol{\theta}\mid \mathbf{D}_E, \Gamma\in (\Gamma_l, \Gamma_u))
\end{equation}
where, the posterior distribution $\pi(\boldsymbol{\theta}\mid \mathbf{D}_E)$ is proportional to the likelihood followed from the analysis model in (\ref{eq:AM}) and a baseline design prior $\pi_d(\boldsymbol{\theta})$ that can be non-informative in absence of information besides $\mathbf{D}_E$. The Bayesian type I error probability and power are then given as,
\begin{equation}
\int\int P(\tau(\mathbf{D})>\nu\mid \boldsymbol{\theta})\pi_d^j(\boldsymbol{\theta})d\boldsymbol{\theta}d\mathbf{D}_I, \hskip 20pt j = 0, 1.
\end{equation}
The design parameters, including the decision boundary $\nu$ and the size of $\mathbf{D}_I$ ($N_I$) are specified to control the Bayesian type I error probability and achieve a desired level of Bayesian power. 

\cite{psioda_bayesian_2019} proposed Monte Carlo integration for calculating the Bayesian operating characteristics and sample size determination. To reduce the computational cost they proposed to use a large sample normal approximation for the posterior to obtain $\tau(\mathbf{D})$. However, precise approximations of these criteria require a prohibitively large number of iterations due to the multiple integral in multi-parameter models. \cite{Golchi2025BARTBvM} proposed a procedure for efficient estimation of Bayesian operating characteristics in presence of nuisance parameters that also relies on the large sample normality of the posterior but uses predictive modeling to further reduce the computational burden. This approach which relies on an analytic power function derived based on the approximate normality of the posterior whose variance is predicted over the parameter space based on a relatively small number of draws from the design prior. In Section~\ref{sec:app} we use this procedure to generate Bayesian power curves and perform sample size determination.

\section{Simulations}
\label{sec:sim}
We conduct a simulation study to assess the analysis framework under various scenarios. We define the data generating mechanism to closely mimic the data structure in the motivating example presented in Section~\ref{Sec:me}. 
Data are generated for an internal RCT and two external data sources, a retrospective cohort study (RES) where all patients were treated under the control arm and a single arm clinical trial (SCT) where all patients received the treatment. The data generating model is as follows, 
\begin{equation*}
Y_n \sim \text{Binom}\left(N_j, \text{logit}^{-1} \left( \mathbf{x}_n^T\boldsymbol{\beta} +(\mathbf{s}_n^T\boldsymbol{\psi})a_n\right)\right), \hskip 20pt   j = 0, 1, 2; \hskip 5pt n = 1, \ldots, N_j.
\end{equation*}
where $\mathbf{x}=(1, x_1, x_2, x_3, x_4)$ is a vector including two continuous and two binary prognostic covariates where $x_4$ is assumed to be a predictive variable (effect modifier) as well, i.e., $\mathbf{s}=(1, x_4)$. Therefore, $\boldsymbol{\beta}$ is a vector of size five and $\boldsymbol{\psi} = (\psi_0, \psi_1)$. For the RES, $a_n=0$, for $n= 1, \ldots, N_1$, for the SCT, $a_n = 1$, for $n = 1, \ldots, N_2$, and for the RCT, $a_n \sim \text{Bernoulli(0.5)}$, $n = 1, \ldots, N_0$ to reflect balanced randomization. 

The main effect parameters $\boldsymbol{\beta}$ and $\psi_0$ are held fixed for all data sources and across all simulation scenarios. 
The simulation scenarios are defined to assess the sensitivity of the estimation accuracy and precision to the level of discordance between external and internal data, the sample size and correlation between measured and ``unmeasured" prognostic covariates. We induce discrepancies between the external and internal data with respect to the covariate distributions and the treatment-covariate interaction parameter, $\psi_1$, that drives effect heterogeneity. The two continuous covariates are generated from a multivariate normal distribution with a correlation parameter $\rho$ that is set to zero in half of the simulation scenarios. Table~\ref{tab:sim2scn} summarizes the simulation scenarios. Details regarding the specific distributions and parameter values are provided in Table S2 in the Supplementary Material.

Similarly, we define a set of simulation scenarios under the null hypothesis of no treatment effect in either subgroup to generate the RCT. These simulation scenarios are presented in Table~\ref{tab:simNull}. The first row represents the case that there is discordance in internal and external covariate distributions but the SCT data are also generated with a null treatment effect. The other two scenarios assume a small and moderate effect for the SCT which is expected to induce bias and inflate the type I error rate when external data are borrowed.

Each generated data set in the simulation study is analyzed using the individually weighted model introduced in (\ref{eq:wIPD}) with the likelihood given by a model of similar form as the data generating model but mis-specified with respect to inclusion of prognostic variables by excluding $x_2$. We consider four versions of the individually weighted model, the truncated and untruncated weights including all prognostic-only variables, and truncated and untruncated weights with the same set of covariates as in the analysis model (i.e., all but $x_2$). The prior distributions are specified as weakly informative default priors, $N(0, 2.5)$. The individually weighted model is implemented using cmdStan. In addition, each data set is analyzed by LEAP with a logistic regression and the same set of variables and default priors implemented in the hdbayes R package.

We define the target of inference as the marginal treatment effect among patients with $\mathbf{s} = (1,1)^T$ obtained by marginalizing over prognostic-only variables, 
\begin{equation*}
\Gamma_s = \iint \left[E(\mathbf{Y}\mid x_1, x_3, x_4=1, A=1;\boldsymbol{\theta}) - E(\mathbf{Y}\mid x_1, x_3, x_4=1, A=0;\boldsymbol{\theta})\right]  p(x_1,x_3)dx_1dx_3
\end{equation*}
As this estimand depends not only on the model parameters but also on the covariate distributions its true value varies across different patient populations. Therefore, an analysis that incorporates external, non-randomized data and fails to adjust for confounders is expected to generate biased estimates where bias may arise not only from parameters but also mis-characterized covariate distributions. 

Given samples, $\boldsymbol{\theta}_r = (\beta_{r,0}, \beta_{r,1}, \beta_{r,2}, \beta_{r,3}, \psi_{r,0}, \psi_{r,1})$, $r = 1,\ldots, R$ from the joint posterior distribution, draws from the posterior distribution of the marginal estimand are obtained using Bayesian bootstrap \cite{} as follows:
\begin{align*}
\Gamma_{r,s} = \sum_{n=1}^N \nu_{r,n}&[\text{logit}^{-1}\left( \beta_{r,0} + \beta_{r,1}x_{n,1} + \beta_{r,2}x_{n,3} + \beta_{r,3} + \psi_{r,0} + \psi_{r,1}\right)\\ & - \text{logit}^{-1}\left(\beta_{r,0} + \beta_{r,1}x_{n,1} + \beta_{r,2}x_{n,3} + \beta_{r,3} \right)]
\end{align*}
where $\nu_{r,n}$ are Dirichlet weights generated from $\text{Dir}(\mathbf{1}_N)$.

\begin{figure}
	\begin{center}
		\includegraphics[width= \textwidth]{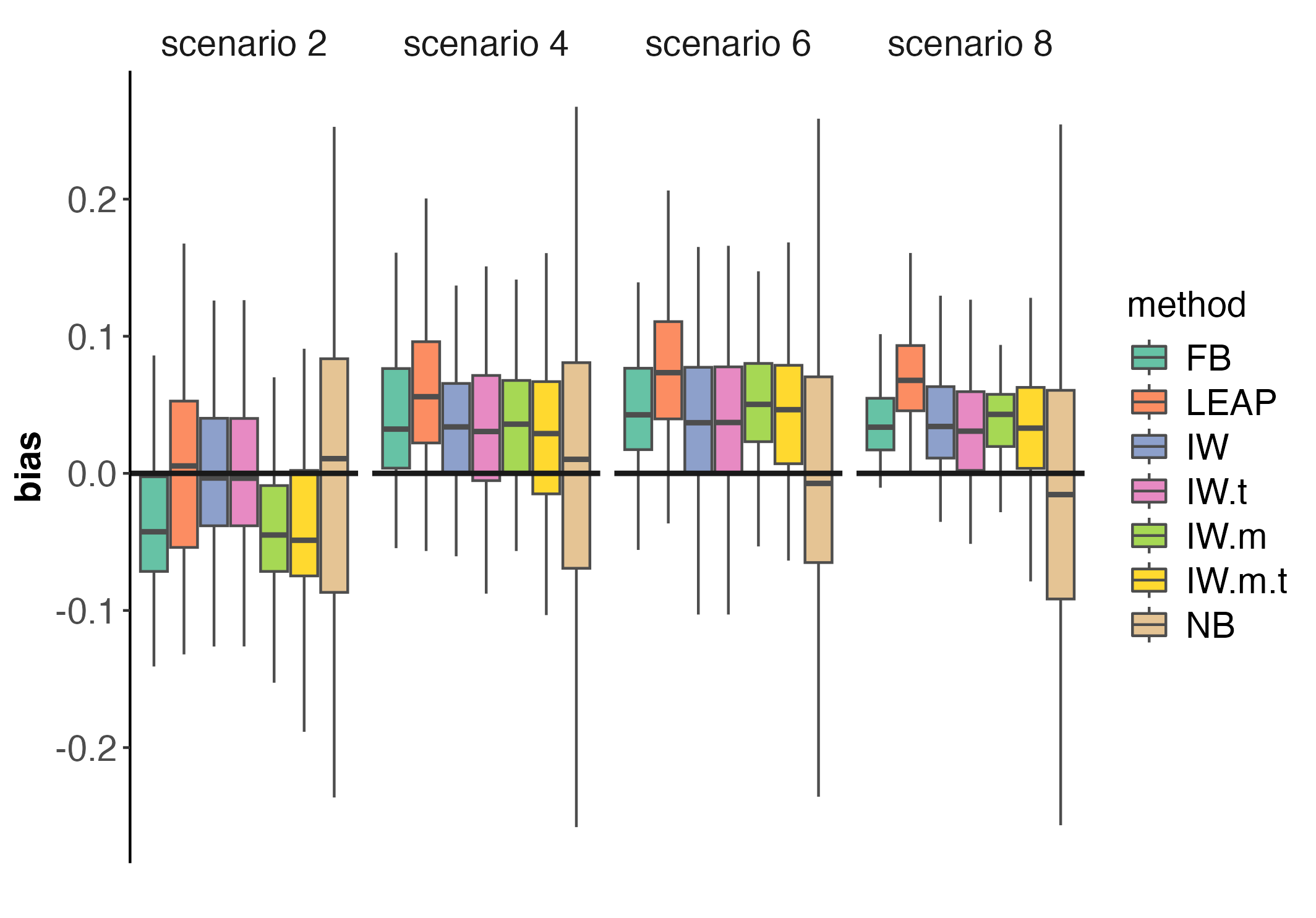}
		\caption{Deviation of posterior medians from the true marginal subgroup effect across 100 Monte Carlo iterations}
		\label{fig:sim2bias}
        \end{center}
\end{figure}

\begin{figure}	
\begin{center}
		\includegraphics[width= \textwidth]{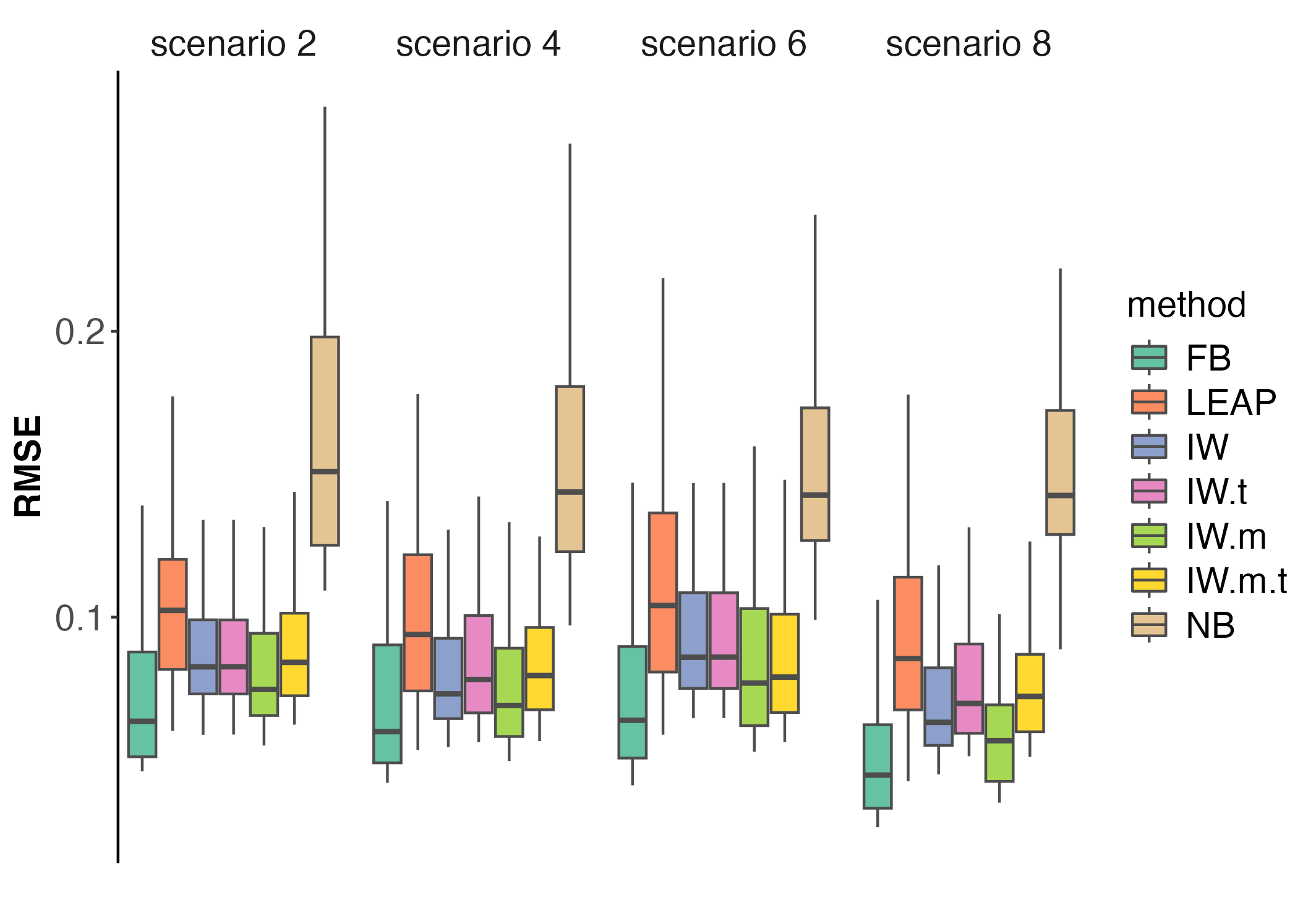}
		\caption{Posterior root mean squared error fore the marginal subgroup effect across 100 Monte Carlo iterations}
		\label{fig:sim2rmse}
\end{center}
\end{figure}

The results are provided in Figures~\ref{fig:sim2bias} and~\ref{fig:sim2rmse}, and Table~\ref{tab:sim2res}. As the results indicated little sensitivity to the correlation between the two continuous covariates, we only present the correlated scenarios in Figures~\ref{fig:sim2bias} and~\ref{fig:sim2rmse} and uncorrelated ones in Table~\ref{tab:sim2res}. 

Figure~\ref{fig:sim2bias} shows the distribution of the deviation of posterior median from the true value of the marginalized subgroup effect for the seven analysis models. As expected, an analysis that leverages no external data is the only unbiased analysis across all scenarios while it results in the largest estimation variance. On the opposite extreme, full borrowing results in bias systematically across scenarios while maintaining the smallest estimation variance due to the large sample size. Under scenarios 1 and 2 where the external and internal data are discordant only with respect to the covariates, LEAP and the individually weighted model IW and its truncated version IW.t where the weights are obtained with the full set of covariates achieve unbiased estimates with IW and IW.t attaining lower variance compared to LEAP. The weighted models with a missing covariate, IW.m and IW.m.t fail to fully adjust for the discordance and result in bias. Under scenarios 3-8 all borrowing models lead to bias. Curiously, LEAP produces the largest bias by a small margin compared to other models including the full borrowing. This is more emphasized under scenarios 7 and 8 where there is discordance between the data sources both with respect to covariate distributions and parameters, and the external data size is large. Also, as anticipated, truncation of the weights only makes a difference when the external sample size is large.

Table~\ref{tab:simNull_res} presents the type I error rate estimates with two decision thresholds of $\nu = 0.95$ and $\nu = 0.975$. For all methods the type I error rate increases as the level of discordance between internal and external data increases. When discordance is only in the covariate distributions the IW and IW.t control the type I error rate while LEAP, IW.m and IW.m.t result in some inflation. As discordance with respect to the underlying effect parameter is introduced, all borrowing methods result in some type I error rate inflation with the IW and IW.t maintaining the smallest levels.

\begin{table}
\centering
\begin{tabular}{c|c|c|c|c}
\hline
Scenario & Discordance in & Discordance in $\psi_1$ & $\rho$ & $(N_1, N_2)$\\
 & covariate distributions &  &  & \\
\hline
1 &\ding{51}&\ding{55}&0& (500,100)\\
2 &\ding{51}&\ding{55}&0.5&(500,100)\\
3 &\ding{55}&\ding{51}&0&(500,100)\\
4 &\ding{55}&\ding{51}&0.5&(500,100)\\
5 &\ding{51}&\ding{51}&0&(500,100)\\
6 &\ding{51}&\ding{51}&0.5&(500,100)\\
7 &\ding{51}&\ding{51}&0&(1000,300)\\
8 &\ding{51}&\ding{51}&0.5&(1000,300)\\
\hline
  \end{tabular} 
   \caption{Summary of simulation scenarios under the alternative hypothesis}
   \label{tab:sim2scn}
   \end{table}

\begin{table}
 \centering 
\begin{tabular}{c|c|c|c}
\hline
Scenario & $\psi_1^{\text{SCT}}$ & $\rho$ & $(N_1, N_2)$\\
 &   &  & \\
\hline
01 &0&0& (500,100)\\
02 &0.3&0.5&(500,100)\\
03 &0.6 &0.5&(500,100)\\
\hline
  \end{tabular} 
   \caption{Summary of simulation scenarios under the null hypothesis; there is discordance in covariate distributions under all scenarios.}
   \label{tab:simNull}
   \end{table}

\begin{table}
    \centering 
\begin{tabular}{ l|c|c|c|c|c|c|c|c }
\hline
&\multicolumn{4}{|c|}{scenario 1} &\multicolumn{4}{|c|}{scenario 3} \\
\hline
 method & coverage & bias & rmse & power & coverage & bias & rmse & power \\
\hline
 FB & 0.77 & -0.048 & 0.080 & 0.44 & 0.90 & 0.033 & 0.073 & 0.92\\
 LEAP & 0.97 & -0.010 & 0.111 & 0.43& 0.90 & 0.047 & 0.114 & 0.71\\

 IW & 1.00 & -0.005 & 0.090 & 0.44 & 0.92 & 0.028 & 0.092 & 0.80 \\

 IW.t & 1.00 & -0.005 & 0.090 & 0.44& 0.90 & 0.025 & 0.098 & 0.75\\

 IW.m & 0.88 & -0.049 & 0.091 & 0.26 & 0.92 & 0.029 & 0.084 & 0.87 \\

 IW.m.t & 0.89 & -0.049 & 0.100 & 0.25& 0.92 & 0.023 & 0.100 & 0.67\\

 NB & 0.94 & 0.012 & 0.169 & 0.30& 0.92 & -0.003 & 0.175 & 0.21\\
\hline
& \multicolumn{4}{|c|}{scenario 5}& \multicolumn{4}{|c|}{scenario 7}\\
\hline
 FB & 0.93 & 0.026 & 0.066 & 0.96 & 0.71 & 0.036 & 0.054 & 1.00\\
 LEAP &  0.89 & 0.051 & 0.101 & 0.88 & 0.70 & 0.065 & 0.099 & 0.91\\

 IW &  0.99 & 0.016 & 0.092 & 0.61& 0.97 & 0.028 & 0.069 & 0.92\\

 IW.t &  0.99 & 0.016 & 0.092 & 0.61 & 0.97 & 0.027 & 0.077 & 0.86\\

 IW.m &  0.95 & 0.025 & 0.078 & 0.89 & 0.83 & 0.037 & 0.062 & 1.00\\

 IW.m.t &  0.92 & 0.025 & 0.086 & 0.83& 0.89 & 0.040 & 0.089 & 0.87\\

 NB &  0.92 & -0.001 & 0.171 & 0.23& 0.97 & -0.018 & 0.173 & 0.22\\
\hline
\end{tabular}
\caption{Simulation study results for scenarios 1, 3, 5, and 7 under the alternative hypothesis. FB: full borrowing, LEAP: latent exchangeability model, IW: individually weighted, IW.t: individually weighted \& truncated, IW.m: individually weighted with missing covariate, IW.m.t: individually weighted with missing covariate \& truncated, NB: no borrowing.}
    \label{tab:sim2res}
\end{table}

\begin{table}
    \centering 
\begin{tabular}{ l|c|c|c|c|c|c|c|c|c}
\hline
&\multicolumn{2}{|c|}{scenario 01 } & \multicolumn{2}{|c|}{scenario 02 }&\multicolumn{2}{|c|}{scenario 03 }\\
\hline
method&$\nu = 0.95$ & $\nu = 0.975$ &$\nu = 0.95$ & $\nu = 0.975$ & $\nu = 0.95$ & $\nu = 0.975$ \\
\hline
 FB & 0.14 & 0.10 & 0.48 & 0.35 & 0.85 &0.79\\
 LEAP &  0.14 & 0.09 & 0.29 & 0.19 & 0.58 & 0.48\\

 IW &  0.03 & 0.02 & 0.12 & 0.06 & 0.35 & 0.22\\

 IW.t &  0.03 & 0.02 & 0.12 & 0.06 & 0.35 & 0.22\\

 IW.m &  0.13 & 0.08 & 0.32 & 0.17& 0.70 & 0.57\\

 IW.m.t &  0.11 & 0.07 & 0.21 & 0.15& 0.58 & 0.43\\

 NB &  0.06 & 0.03 & 0.02 & 0.01 & 0.05 &0.04\\
\hline
\end{tabular}
\caption{Type I error rate with two decision boundaries. FB: full borrowing, LEAP: latent exchangeability model, IW: individually weighted, IW.t: individually weighted \& truncated, IW.m: individually weighted with missing covariate, IW.m.t: individually weighted with missing covariate \& truncated, NB: no borrowing.}
    \label{tab:simNull_res}
\end{table}

\section{Application to the XParTS-II trial}
\label{sec:app}
We consider the XParTS-II trial and its accompanying external data sources that were introduced in Section~\ref{Sec:me} from analysis and design perspectives. 

{\bf Analysis} Focusing on the tumor response as the outcome and the marginalized treatment effect in the subgroup of patients with recurrent gastric cancer, we first present a re-analysis of the trial data with no borrowing, individually weighted borrowing and full borrowing. 

\begin{figure}

		\centering
		\includegraphics[width= \textwidth]{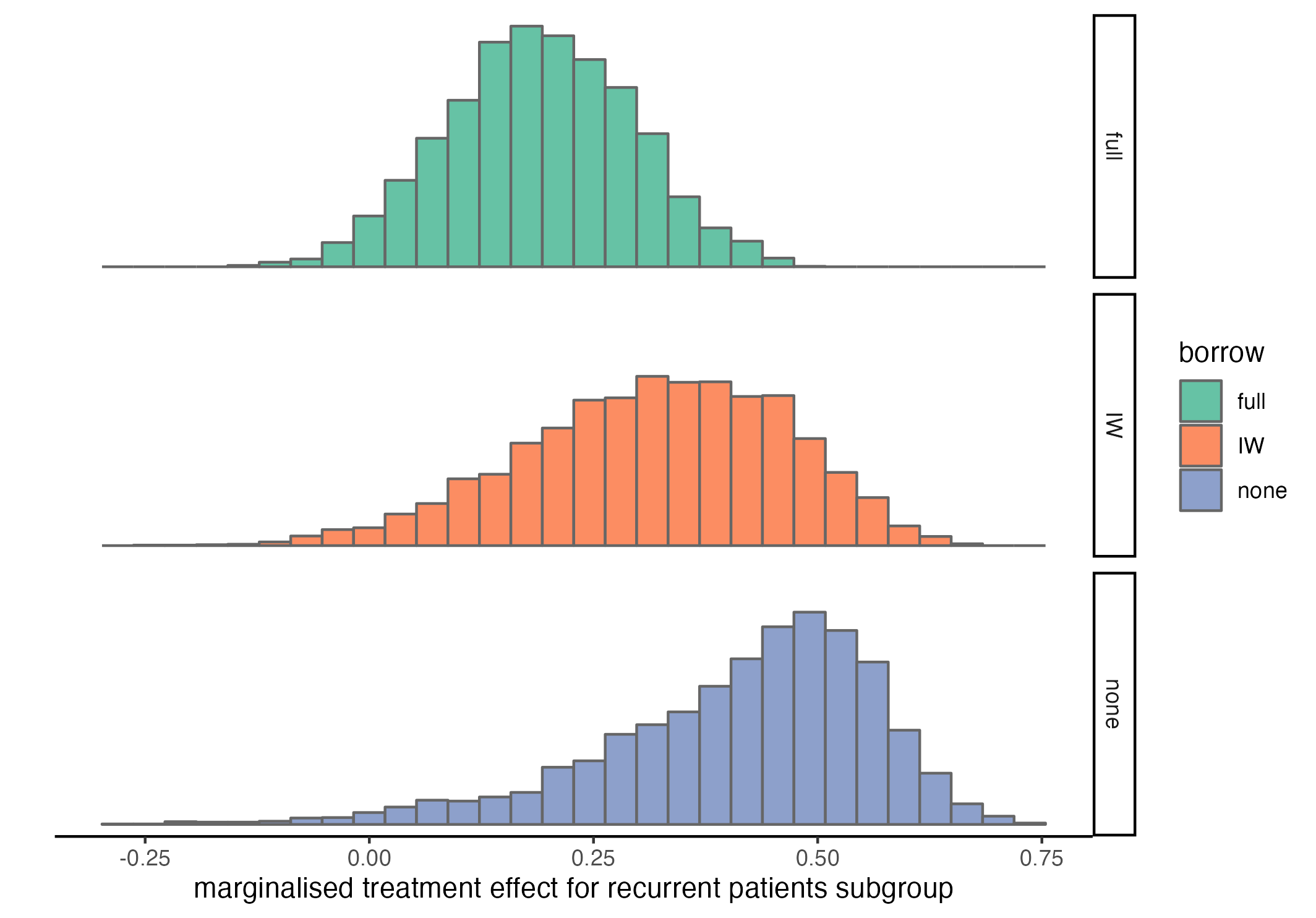}
	
	\caption{The posterior distributions of the marginalized treatment effect among patients with recurrent disease status based on only the XParTS-II trial data (bottom), partial borrowing (middle) and full borrowing (top).}
	\label{fig:DA_res}
\end{figure}

For calculating the individual weights we use the complete set of covariates available across the data sets including sex, age, site of metastasis (three binary variables indicating the site as lymph node, peritoneum or liver), and performance status. For the analysis we model the binary tumor response using a logistic regression that includes the disease status, recurrent or metastatic, as both a prognostic and predictive covariate.
\begin{equation*}
Y \sim \text{Binom}\left(N, \text{logit}^{-1} \left( \beta_0 + \beta_1x +(\psi_0 + \psi_1x)a\right)\right),
\end{equation*}
The analysis results are presented in Figure~\ref{fig:DA_res} as the posterior distribution of the marginalized treatment effect among patients with recurrent gastric cancer based on only the XParTS-II trial data, using a naïve analysis that combines the data from XParTS-II trial with XParTS-I and the retrospective study as well as the individually weighted model. The posterior distribution resulting from the analysis of XParTS-II trial data alone (bottom histogram) suggests a stronger effect in this subgroup although the posterior variance is rather large due to the small subgroup size (only 25 patients out of the 110). In fact, the long left tail of the posterior distribution indicates that smaller effect sizes remain plausible. Integrating the external data fully into the analysis (top histogram) further supports a smaller effect. The individually weighted analysis that adjusts for baseline covariate similarities (middle histogram) results in a compromise between the two other analyses with respect to the magnitude of the estimated effect and the uncertainty. However, partial borrowing results in a very modest reduction in uncertainty as the external data sample sizes are small.

{\bf Design} From a design perspective, we consider the question of how the XParTS-II trial could have been designed with the objective of investigating effect heterogeneity between the two subgroups and estimating the marginalized treatment effect among patients with recurrent disease status while leveraging the available XParTS-I and retrospective data. Since the external data only include patients with recurrent disease status, the individual coefficients ($\beta_0$, $\beta_1$, $\psi_0$, and $\psi_1$) are not identifiable in an analysis of the external data alone. These data provide information about $\beta_0+\beta1$ as well as $\psi_0 + \psi_1$ which can be leveraged toward the design. However, effect heterogeneity can be only inferred once data on the metastatic patient subgroup becomes available.

\begin{figure}

		\centering
		\includegraphics[width= \textwidth]{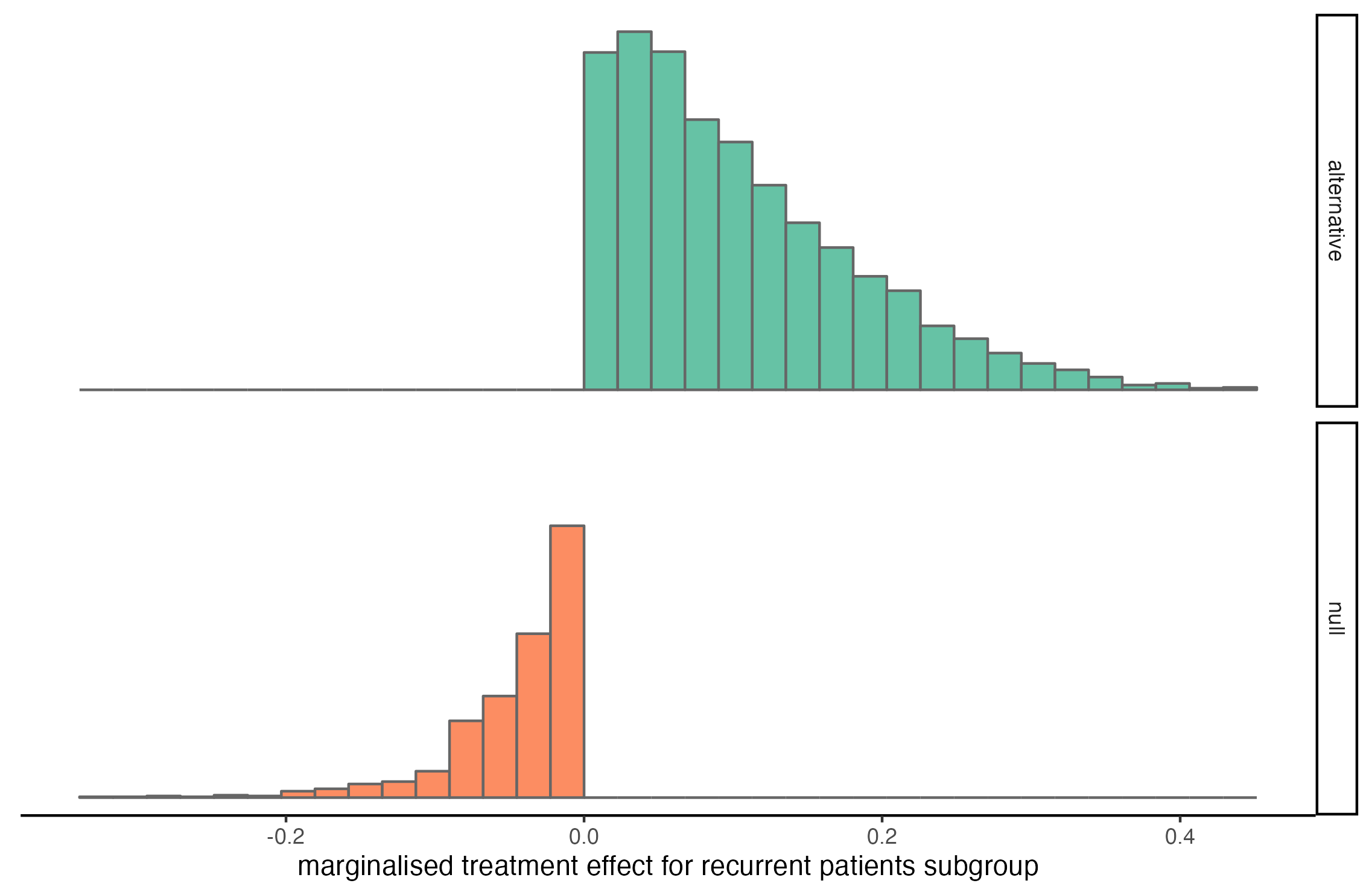}
	
	\caption{The posterior distributions of the marginalized treatment effect given the external data used to define the null and alternative design priors}
	\label{fig:design_prior}
\end{figure}

As the operating characteristics we use the Bayesian type I error rate and power as defined in Section~\ref{sec:des} with respect to default null and alternative design priors obtained from the posterior distribution of the model parameters given the external data. The default null and alternative design priors are defined as the joint distribution of model parameters that result in $\Gamma\leq0$ and $\Gamma>0$, respectively. Figure~\ref{fig:design_prior} shows these design priors for $\Gamma$. 

We define the subgroup efficacy decision as $P(\Gamma>0\mid \mathbf{D}_I, \mathbf{D}_E)>\nu$, and specify $\nu$ as a probability threshold that maintains the Bayesian type I error rate below 0.1. As extensively discussed in the literature \citep{psioda_bayesian_2019, koppschneider_power_2020}, inflation of type I error rate when borrowing on the treatment effect is unavoidable and it is reasonable to assume a higher tolerated type I error rate compared to the typical 0.05 in this case. We will then proceed with obtaining the Bayesian power curves for the XParTS-II trial when analyzed alone or with the external data to inform sample size determination. 

Considering the size of external data and the modest effect size (Figure~\ref{fig:design_prior}), it is trivial to see that the redesign of the XParTS-II to target the subgroup effect would be a futile exercise as the required sample sizes are far larger than the size of all three data sets. Therefore, for illustrative purposes, we perform this hypothetical re-design exercise by synthesizing larger external data ($N_E=1000$) via resampling from the real data to maintain the same proportion of control to treatment and covariate distributions as in the real data.

\begin{figure}

		\centering
		\includegraphics[width= \textwidth]{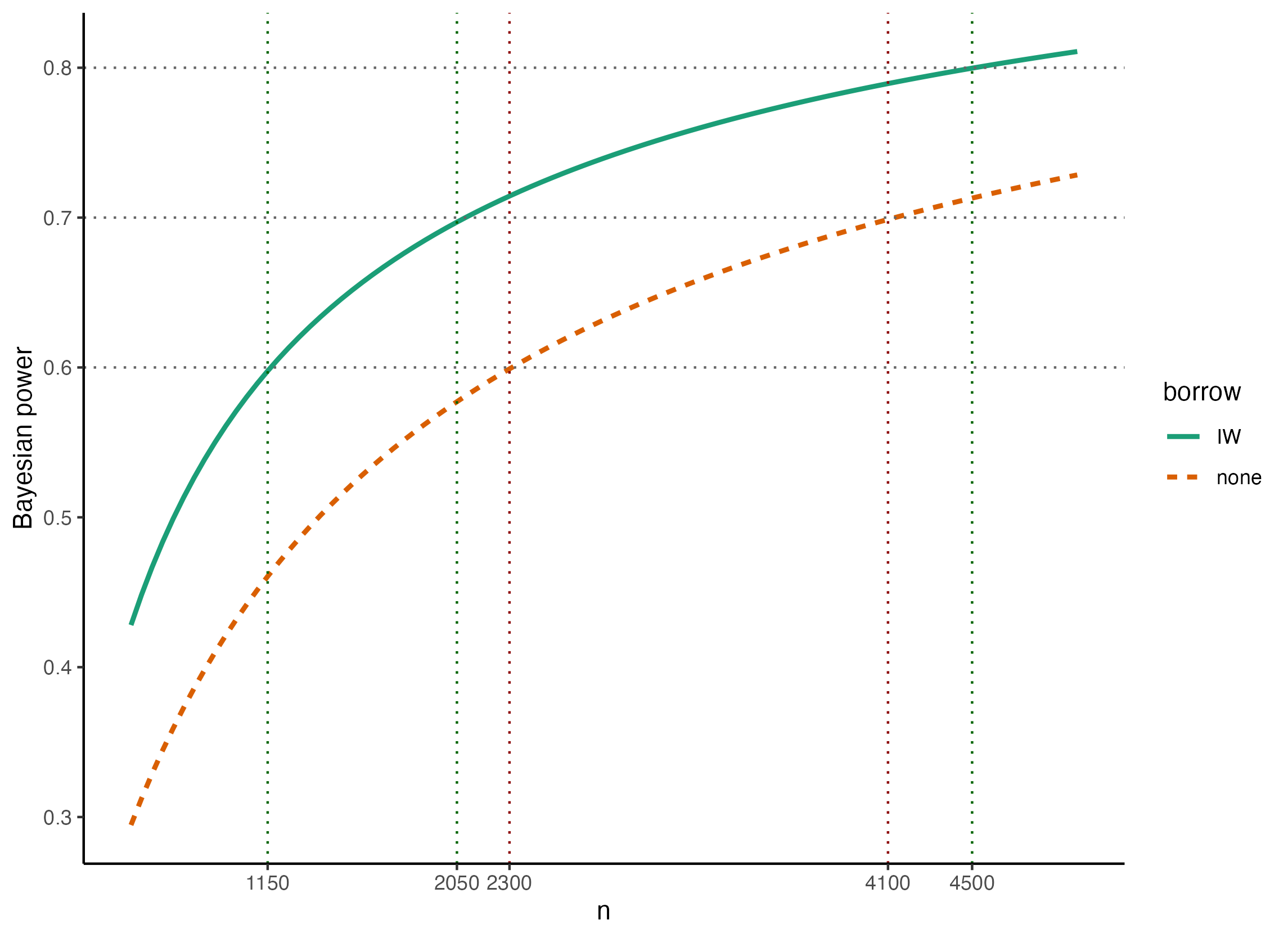}
	
	\caption{Bayesian power curves with and without borrowing. The dotted vertical lines indicate the sample size required to achieve 60\%, 70\% and 80\% power.}
	\label{fig:ssd}
\end{figure}

Figure~\ref{fig:ssd} shows the Bayesian power curves against the sample size with and without borrowing. The results indicate that, without incorporating the external data, a trial twice as large would be required to achieve the same level of power. 

\section{Discussion}
\label{sec:dis}

In this paper we considered information borrowing toward the design and analysis of clinical trials with the objective of detecting effect heterogeneity and establishing subgroup specific effects. As the analysis prior we proposed an individually weighted likelihood of the external data where weights are specified based on the fit of each individual to the target population based on a set of covariates that are at least partially available across the internal and external data sources. We used the posterior predictive similarity function as a measure of fit. In cases where the external data sizes are large compared to that of the internal data, we recommend truncating the weights distribution to achieve approximate balance between the external and internal sample sizes.

Our approach is a static partial borrowing method which only uses covariate data to adjust the level of information borrowed from external sources.  We acknowledge that the proposed model is subject to the criticisms generally made to static borrowing approaches and those that only use covariate and not outcomes to adjust the borrowing level. In our simulation study we assessed the performance under various scenarios defined based on the level and nature of discordance between internal and external data and made comparisons with the LEAP which is the closest dynamic borrowing equivalent. Under the considered scenarios our model achieved comparable or better results in terms of accuracy and precision compared to LEAP. While we admit that our approach lacks the flexibility and elegance of fully Bayesian methods like LEAP, we emphasize the simplicity of the approach which is considered an advantage from a practical perspective.


In addition to the analysis model, we offered a Bayesian design framework where partially informative external data may be used to construct design priors. We illustrated the design procedure in the context of the redesign of the XParTS-II trial. While we presented a sample size determination exercise for a fixed design, we note that the design framework can be seamlessly generalized to adaptive designs, particularly, adaptive enrichment designs.

To conclude, the present manuscript provides a practical framework for utilizing external data in design and analysis of precision trials. The presented information borrowing method is simple and straightforward to implement, yet it achieves comparable performance to top-performing existing methods. In addition, we highlight the important role of external data, albeit incomplete, in the design of precision trials.




\begin{thebibliography}{}
	
	\bibitem[\protect\citename{Alt {\em et~al.}, }2024]{alt_leap_2024}
	Alt, Ethan~M, Chang, Xiuya, Jiang, Xun, Liu, Qing, Mo, May, Xia, Hong~Amy, \& Ibrahim, Joseph~G. 2024.
	\newblock {LEAP}: the latent exchangeability prior for borrowing information from historical data.
	\newblock {\em Biometrics}, {\bf 80}(3), ujae083.
	\newblock \_eprint: https://academic.oup.com/biometrics/article-pdf/80/3/ujae083/59365947/ujae083.pdf.
	
	\bibitem[\protect\citename{Best {\em et~al.}, }2025]{best_beyond_2025}
	Best, Nicky, ~, Maxine, Ajimi, ~, Beat, Neuenschwander, ~, Gaëlle, Saint-Hilary, , \& Wandel, Simon. 2025.
	\newblock Beyond the {Classical} {Type} {I} {Error}: {Bayesian} {Metrics} for {Bayesian} {Designs} {Using} {Informative} {Priors}.
	\newblock {\em Statistics in Biopharmaceutical Research}, {\bf 17}(2), 17(2), 183–196.
	\newblock Publisher: ASA Website \_eprint: https://doi.org/10.1080/19466315.2024.2342817.
	
	\bibitem[\protect\citename{Carvalho \& Ibrahim, }2021]{carvalho2021normalized}
	Carvalho, Luiz~Max, \& Ibrahim, Joseph~G. 2021.
	\newblock On the normalized power prior.
	\newblock {\em Statistics in Medicine}, {\bf 40}(24), 5251--5275.
	
	\bibitem[\protect\citename{Chen \& Ibrahim, }2000]{CheIbr2000}
	Chen, M.H., \& Ibrahim, J.G. 2000.
	\newblock Power prior distributions for regression models.
	\newblock {\em Statistical Science}, {\bf 15}(1), 46--60.
	
	\bibitem[\protect\citename{Chen {\em et~al.}, }2016]{CheCheMor16}
	Chen, Yiyi, Chen, Zunqiu, \& Mori, Motomi. 2016.
	\newblock {A new statistical decision rule for single-arm phase II oncology trials}.
	\newblock {\em Statistical Methods in Medical Research}, {\bf 25}(1), 118--132.
	
	\bibitem[\protect\citename{Daniells {\em et~al.}, }2025]{daniells_incorporating_2025}
	Daniells, Libby, Mozgunov, Pavel, Barnett, Helen, Bedding, Alun, \& Jaki, Thomas. 2025.
	\newblock Incorporating historic information to further improve power when conducting {Bayesian} information borrowing in basket trials.
	\newblock {\em Biostatistics}, {\bf 26}(1), kxaf016.
	
	\bibitem[\protect\citename{Dempster {\em et~al.}, }1982]{DemSelWee1981}
	Dempster, A.~P., Selwyn, M.~R., \& Weeks, B.~J. 1982.
	\newblock Combining Historical and Randomized Controls for Assessing Trends in Proportions.
	\newblock {\em Journal of the American Statistical Association}, {\bf 78}(382), 221--227.
	
	\bibitem[\protect\citename{Duan, }2005]{duan2005modified}
	Duan, Yuyan. 2005.
	\newblock {\em A modified Bayesian power prior approach with applications in water quality evaluation}.
	\newblock Ph.D. thesis, Virginia Polytechnic Institute and State University.
	
	\bibitem[\protect\citename{Duan {\em et~al.}, }2006]{duan2006evaluating}
	Duan, Yuyan, Ye, Keying, \& Smith, Eric~P. 2006.
	\newblock Evaluating water quality using power priors to incorporate historical information.
	\newblock {\em Environmetrics: The Official Journal of the International Environmetrics Society}, {\bf 17}(1), 95--106.
	
	\bibitem[\protect\citename{Gelfand \& Wang, }2002]{gelfand_simulation-based_2002}
	Gelfand, Alan~E., \& Wang, Fei. 2002.
	\newblock A simulation-based approach to {Bayesian} sample size determination for performance under a given model and for separating models.
	\newblock {\em Statistical Science}, {\bf 17}(2), 193--208.
	\newblock Publisher: Institute of Mathematical Statistics.
	
	\bibitem[\protect\citename{Golchi \& Hagar, }2025]{Golchi2025BARTBvM}
	Golchi, Shirin, \& Hagar, Luke. 2025.
	\newblock Bayesian Design of Experiments in the Presence of Nuisance Parameters.
	\newblock {\em arXiv preprint arXiv:2508.03948}.
	
	\bibitem[\protect\citename{Gower, }1971]{Gower1971}
	Gower, J.~C. 1971.
	\newblock A General Coefficient of Similarity and Some of Its Properties.
	\newblock {\em Biometrics}, {\bf 27}(4), 857--871.
	
	\bibitem[\protect\citename{Hobbs {\em et~al.}, }2011]{hobbs_hierarchical_2011}
	Hobbs, Brian~P., Carlin, Bradley~P., Mandrekar, Sumithra~J., \& Sargent, Daniel~J. 2011.
	\newblock Hierarchical {Commensurate} and {Power} {Prior} {Models} for {Adaptive} {Incorporation} of {Historical} {Information} in {Clinical} {Trials}.
	\newblock {\em Biometrics}, {\bf 67}(3), 1047--1056.
	
	\bibitem[\protect\citename{Ibrahim {\em et~al.}, }1998]{IbrRyaChe1998}
	Ibrahim, J.~G., Ryan, L.~M., \& Chen, M.~H. 1998.
	\newblock Using Historical Controls to Adjust for Covariates in Trend Tests for Binary Data.
	\newblock {\em Journal of the American Statistical Association}, {\bf 93}(444), 1282--1293.
	
	\bibitem[\protect\citename{Ibrahim {\em et~al.}, }2015]{IbrCheGwo15}
	Ibrahim, J.~G., Chen, M.~H., Gwon, Y., \& Chen, F. 2015.
	\newblock The Power Prior: Theory and Applications.
	\newblock {\em Statistics in Medicine}, {\bf 34}(28), 3724--3749.
	
	\bibitem[\protect\citename{Jiang {\em et~al.}, }2023]{jiang_elastic_2023}
	Jiang, Liyun, Nie, Lei, \& Yuan, Ying. 2023.
	\newblock Elastic priors to dynamically borrow information from historical data in clinical trials.
	\newblock {\em Biometrics}, {\bf 79}(1), 49--60.
	
	\bibitem[\protect\citename{Kopp‐Schneider {\em et~al.}, }2020]{koppschneider_power_2020}
	Kopp‐Schneider, Annette, Calderazzo, Silvia, \& Wiesenfarth, Manuel. 2020.
	\newblock Power gains by using external information in clinical trials are typically not possible when requiring strict type {I} error control.
	\newblock {\em Biometrical Journal. Biometrische Zeitschrift}, {\bf 62}(2), 361--374.
	
	\bibitem[\protect\citename{Kwiatkowski {\em et~al.}, }2024]{kwiatkowski_case_2024}
	Kwiatkowski, Evan, Zhu, Jiawen, Li, Xiao, Pang, Herbert, Lieberman, Grazyna, \& Psioda, Matthew~A. 2024.
	\newblock Case weighted power priors for hybrid control analyses with time-to-event data.
	\newblock {\em Biometrics}, {\bf 80}(2), ujae019.
	\newblock \_eprint: https://academic.oup.com/biometrics/article-pdf/80/2/ujae019/57099861/ujae019.pdf.
	
	\bibitem[\protect\citename{Lewis {\em et~al.}, }2019]{LewSarZhu19}
	Lewis, C.J., Sarkar, S., Zhu, J., \& Carlin, B.P. 2019.
	\newblock Borrowing from Historical Control Data in Cancer Drug Development: A Cautionary Tale and Practical Guidelines.
	\newblock {\em Statistics in Biopharmaceutical Research}, {\bf 11}(1), 67--78.
	
	\bibitem[\protect\citename{Lu {\em et~al.}, }2022]{Lu02012022}
	Lu, Nelson, Wang, Chenguang, Chen, Wei-Chen, Li, Heng, Song, Changhong, Tiwari, Ram, Xu, Yunling, \& Yue, Lilly~Q. 2022.
	\newblock Propensity score-integrated power prior approach for augmenting the control arm of a randomized controlled trial by incorporating multiple external data sources.
	\newblock {\em Journal of Biopharmaceutical Statistics}, {\bf 32}(1), 158--169.
	\newblock PMID: 34756158.
	
	\bibitem[\protect\citename{Müller {\em et~al.}, }2011]{Muller_ppm_2011}
	Müller, P., Quintana, F., \& Rosner, G.L. 2011.
	\newblock A Product Partition Model With Regression on Covariates.
	\newblock {\em Journal of Computational and Graphical Statistics}, {\bf 20}(1), 260--278.
	
	\bibitem[\protect\citename{Neuenschwander {\em et~al.}, }2010]{NeuCapBra2010}
	Neuenschwander, B., Capkun-Niggli, G., Branson, M., \& Spiegelhalter, D.~J. 2010.
	\newblock Summarizing historical information on controls in clinical trials.
	\newblock {\em Clinical Trials}, {\bf 7}(1), 5--18.
	
	\bibitem[\protect\citename{Nishikawa {\em et~al.}, }2018a]{XParTSI}
	Nishikawa, K., Tsuburaya, A., Yoshikawa, T., Takahashi, M., Tanabe, K., Yamaguchi, K., Yoshino, S., Namikawa, T., Aoyama, T., Rino, Y., Kawada, J., Tsuji, A., Taira, K., Kimura, Y., Kodera, Y., Hirashima, Y., Yabusaki, H., Hirabayashi, N., Fujitani, K., Miyashita, Y., Morita, S., \& Sakamoto, J. 2018a.
	\newblock {A phase {II} trial of capecitabine plus cisplatin (XP) for patients with advanced gastric cancer with early relapse after S-1 adjuvant therapy: {XParTS-I} trial.}
	\newblock {\em Gastric Cancer}, {\bf 21}(5), 811--818.
	
	\bibitem[\protect\citename{Nishikawa {\em et~al.}, }2018b]{XParTSII}
	Nishikawa, K., Tsuburaya, A., Yoshikawa, T., Kobayashi, M., Kawada, J., Fukushima, R., Matsui, T., Tanabe, K., Yamaguchi, K., Yoshino, S., Takahashi, M., Hirabayashi, N., Sato, S., Nemoto, H., Rino, Y., Nakajima, J., Aoyama, T., Miyagi, Y., Oriuchi, N., Yamaguchi, K., Miyashita, Y., Morita, S., \& Sakamoto, J. 2018b.
	\newblock {A randomised phase {II} trial of capecitabine plus cisplatin versus S-1 plus cisplatin as a first-line treatment for advanced gastric cancer: Capecitabine plus cisplatin ascertainment versus S-1 plus cisplatin randomised PII trial (XParTS II).}
	\newblock {\em European Journal of Cancer}, {\bf 101}, 220--228.
	
	\bibitem[\protect\citename{O’Hagan \& Stevens, }2001]{ohagan_bayesian_2001}
	O’Hagan, Anthony, \& Stevens, John~W. 2001.
	\newblock Bayesian {Assessment} of {Sample} {Size} for {Clinical} {Trials} of {Cost}-{Effectiveness}.
	\newblock {\em Medical Decision Making}, {\bf 21}(3), 219--230.
	\newblock Publisher: SAGE Publications Inc STM.
	
	\bibitem[\protect\citename{Page \& Quintana, }2018]{PageQuintana2018}
	Page, G.L., \& Quintana, F.A. 2018.
	\newblock Calibrating covariate informed product partition models.
	\newblock {\em Statistics and Computing}, {\bf 28}(4), 1009--1031.
	
	\bibitem[\protect\citename{Psioda \& Ibrahim, }2019]{psioda_bayesian_2019}
	Psioda, Matthew~A, \& Ibrahim, Joseph~G. 2019.
	\newblock Bayesian clinical trial design using historical data that inform the treatment effect.
	\newblock {\em Biostatistics (Oxford, England)}, {\bf 20}(3), 400--415.
	
	\bibitem[\protect\citename{Rover \& Friede, }2019]{RovFri19}
	Rover, C., \& Friede, T. 2019.
	\newblock Dynamically borrowing strength from another study through shrinkage estimation.
	\newblock {\em Statistical Methods in Medical Research}.
	
	\bibitem[\protect\citename{Schmidli {\em et~al.}, }2014]{SchGstRoy14}
	Schmidli, H., Gsteiger, S., Roychoudhury, S., O'Hagan, A., \& Spiegelhalter, D. 2014.
	\newblock Robust meta‐analytic‐predictive priors in clinical trials with historical control information.
	\newblock {\em Biometrics}, {\bf 70}(4), 1023--1032.
	
	\bibitem[\protect\citename{Schwartz {\em et~al.}, }2025]{schwartz_harmonized_2025}
	Schwartz, Daniel, Saha, Riddhiman, Ventz, Steffen, \& Trippa, Lorenzo. 2025.
	\newblock Harmonized estimation of subgroup-specific treatment effects in randomized trials: the use of external control data.
	\newblock {\em Journal of the Royal Statistical Society Series B: Statistical Methodology}, July, qkaf045.
	
	\bibitem[\protect\citename{Shitara {\em et~al.}, }2012]{Retro}
	Shitara, K., Morita, S., Fujitani, K., Kadowaki, S., Takiguchi, N., Hirabayashi, N., Takahashi, M., Takagi, M., Tokunaga, Y., Fukushima, R., Munakata, Y., Nishikawa, K., Takagane, A., Tanaka, T., Sekishita, Y., Sakamoto, J., \& Tsuburaya, A. 2012.
	\newblock {Combination chemotherapy with S-1 plus cisplatin for gastric cancer that recurs after adjuvant chemotherapy with S-1: multi-institutional retrospective analysis.}
	\newblock {\em Gastric Cancer}, {\bf 15}(3), 245--51.
	
	\bibitem[\protect\citename{Tarone, }1981]{Tar1982}
	Tarone, R.~E. 1981.
	\newblock The Use of Historical Control Information in Testing for a Trend in Proportions.
	\newblock {\em Biometrics}, {\bf 38}(1), 215--220.
	
	\bibitem[\protect\citename{Thall \& Simon, }1990]{ThaSim90}
	Thall, Peter~F., \& Simon, Richard. 1990.
	\newblock {Incorporating historical control data in planning phase II clinical trials}.
	\newblock {\em Statistics in Medicine}, {\bf 9}(3), 215--228.
	
	\bibitem[\protect\citename{Weber {\em et~al.}, }2019]{WebLiSea19}
	Weber, S., Li, Y., Seaman, J., Kakizume, T., \& Schmidli, H. 2019.
	\newblock Applying Meta-Analytic Predictive Priors with the R Bayesian evidence synthesis tools.
	\newblock {\em arXiv:1907.00603}.
	
	\bibitem[\protect\citename{Zocholl {\em et~al.}, }2026]{zocholl_informing_2026}
	Zocholl, Dario, Götte, Heiko, Habermehl, Christina, \& Günhan, Burak~Kürsad. 2026.
	\newblock Informing the {Borrowing} {Process} for {Dose}-{Finding} {Trials} by {Estimating} the {Similarity} {Between} {Population}-{Specific} {Dose}-{Toxicity} {Curves}.
	\newblock {\em Pharmaceutical Statistics}, {\bf 25}(1), e70067.
	\newblock \_eprint: https://onlinelibrary.wiley.com/doi/pdf/10.1002/pst.70067.
	
	\bibitem[\protect\citename{Zohar {\em et~al.}, }2008]{ZohTerZho08}
	Zohar, Sarah, Teramukai, Satoshi, \& Zhou, Yinghui. 2008.
	\newblock {Bayesian design and conduct of phase II single-arm clinical trials with binary outcomes: A tutorial}.
	\newblock {\em Contemporary Clinical Trials}, {\bf 29}(4).
	
\end{thebibliography}
\end{document}